\documentclass[12pt,showpacs,showkeys,amsmath,amssymb]{revtex4}
\usepackage{amsmath,amsfonts,amsthm,amscd,amssymb,latexsym}
\usepackage{bm}
\usepackage{dcolumn}
\usepackage{graphicx}
\usepackage{epstopdf}
\usepackage{color}
\usepackage{epsf}
\usepackage{epsfig}
\usepackage{graphicx, epic, eepic, color}

\newcommand{\beq}{\begin{equation}}
\newcommand{\eeq}{\end{equation}}
\begin{document}

\title{Relativistic Tidal Acceleration of Astrophysical Jets}

\author{Donato \surname{Bini}$^{1,2}$}
\author{Carmen \surname{Chicone}$^{3}$}
\author{Bahram \surname{Mashhoon}$^{4}$}

\affiliation{
$^1$Istituto per le Applicazioni del Calcolo ``M. Picone'', CNR, I-00185 Rome, Italy\\
$^2$ICRANet, Piazza della Repubblica 10, I-65122 Pescara, Italy\\
$^3$Department of Mathematics and Department of Physics and Astronomy, University of Missouri, Columbia, Missouri 65211, USA\\
$^4$Department of Physics and Astronomy,
University of Missouri, Columbia, Missouri 65211, USA\\
}

\date{\today}

\begin{abstract}
Within the framework of general relativity, we investigate the tidal acceleration of astrophysical jets relative to the central collapsed configuration (``Kerr source").  To simplify matters, we neglect electromagnetic forces throughout; however, these must be included in a complete analysis.  The rest frame of the Kerr source  is locally defined via the set of hypothetical static observers in the spacetime exterior to the source. Relative to such a fiducial observer fixed on the rotation axis of the Kerr source, jet particles are tidally accelerated to almost the speed of light if their outflow speed is above a certain threshold, given roughly by one half of the Newtonian escape velocity at the location of the reference observer;  otherwise, the particles reach a certain height, reverse direction and fall back toward the gravitational source.  
\end{abstract}


\maketitle

\section{Introduction}

Recent observational data regarding extragalactic jets in M 87 (NGC 4486) and other active galactic nuclei (AGNs) have indicated that relative to the central engine, which is believed to be a rotating  supermassive black hole with an accretion disk, jet acceleration occurs away from the black hole  to Lorentz gamma factors that are much greater than unity~\cite{Algaba:2016yvx,  Lee:2016ctn, Walker:2016cic, Mertens:2016rhi}.  The data are based on the electromagnetic radiation emitted by highly energetic particles propagating in the magnetic field that helps collimate the jets.  These charged particles are subject to electromagnetic as well as gravitational forces. The dynamics of jets would naturally involve both magnetohydrodynamics (MHD) and relativistic gravity~\cite{Felice, BPu}. We concentrate here on the purely gravitational effects  of the central collapsed configuration. In particular, we calculate the contribution of relativistic tidal accelerations to the Lorentz factor of the particles in the outflow away from the central source. The universality of the gravitational interaction implies that our results are independent of the particles' electric charges. 

Imagine a small spherical fluid body either falling radially into, or going away from, a central massive object. In either case the body tends to be elongated along the radial direction due to gravitational tidal effects. Within the context of Einstein's theory of gravitation, such tidal effects have been the subject of a number of previous investigations in connection with astrophysical jets---see~\cite{CM1, CM2, CM3} and the references cited therein.  In~\cite{CM1, CM2, CM3}, the tidal motions of free test particles were studied relative to a fiducial plasma clump in the jet, which was itself considered free as all MHD forces were neglected for the sake of simplicity. Such considerations revealed that particles with relative speeds below a terminal speed $c/\sqrt{2} \approx 0.7\, c$ accelerate toward this speed, while particles with speeds above it decelerate toward it~\cite{CM1, CM2, CM3}. The theoretical approach that was adopted in those treatments was relevant to observational studies of the speeds of jets in galactic microquasars~\cite{Fender, Fender:2014sia}. In the present work, however, the tidal motion is relative to a fiducial observer that is at rest with respect to the central source, as the observational studies indeed refer to the \emph{rest frame} of the gravitational source~\cite{Lee:2016ctn, Algaba:2016yvx,  Walker:2016cic, Mertens:2016rhi}. This circumstance is therefore treated here in detail, as it had not been previously investigated. To simplify our analysis, we neglect MHD and plasma effects that must be included in a complete treatment.

The particles in the jet are affected by the gravitational field of the central engine, which could be characterized by its mass, angular momentum, quadrupole and higher moments. To simplify matters, in the present investigation we consider only the mass $M >0$ and angular momentum $J=Ma\,c >0$ of the central source. That is,  we will henceforth assume the central engine to be a Kerr source characterized by two \emph{independent} parameters $M$ and $J$. To interpret the observational data  \emph{relative} to the Kerr source within the framework of general relativity (GR), one must set up the ensemble of \emph{static} observers in the exterior Kerr spacetime.  Though at rest, these reference observers are accelerated in GR. \emph{This congruence defines locally within GR the rest frame of the gravitational source.} We are particularly interested in static observers along the rotation axis of  the  exterior Kerr spacetime, since it is generally assumed that jets move along the rotation axis of the central collapsed configuration. Moreover, in GR, observables must be spacetime scalars. A natural invariant way to describe the motion of free particles in the jet relative to a nearby static reference observer involves the establishment of a Fermi normal coordinate system~\cite{Synge} in the neighborhood of the static observer and the investigation of geodesic motion in this invariantly defined Fermi coordinate system. 

The striking feature of high-energy bipolar jets in AGNs and microquasars is that matter is repulsed in opposite directions away from the strong gravitational attraction of the central collapsed configuration. The physical processes that are responsible for such repulsion must involve in an essential way plasma effects as well as effects stemming from general relativity (GR). To bring out the role of relativistic gravity in a deeper way, exact solutions of GR have been investigated which contain \emph{cosmic jets}, namely, mathematical constructs that exhibit strong similarities with astrophysical jets~\cite{Chicone:2010aa, Chicone:2010hy, Chicone:2010xr, Chicone:2011ie, Bini:2014esa}. It is important to incorporate such ideas into the theory of astrophysical
 jets~\cite{Bini:2007zzb, JPGM, Tucker:2016wvt, Gariel:2007st, Gariel:2016vql}. The present work is a further contribution in this general direction.

The Kerr metric written in Boyer-Lindquist coordinates $(t,r,\theta,\phi)$ is given by~\cite{Chandra}
\begin{eqnarray}
\label{K1}
-ds^2&=&g_{\alpha\beta}\,dx^\alpha dx^\beta\nonumber\\
&=&-dt^2+\frac{\Sigma}{\Delta}\,dr^2+\Sigma\, d\theta^2 +(r^2+a^2)\sin^2\theta\, d\phi^2\nonumber\\
&& +\frac{2Mr}{\Sigma}\,(dt-a\sin^2\theta\, d\phi)^2\,,
\end{eqnarray}
where 
\beq\label{K2}
\Sigma=r^2+a^2\cos^2\theta\,,\qquad \Delta=r^2-2Mr+a^2\,.
\eeq
If the specific angular momentum of the source $a = J/(Mc)$ is such that $a\le M$, then the source is a Kerr \emph{black hole}.  We use units such that $G= c = 1$, unless specified otherwise. Greek indices run from $0$ to $3$, while Latin indices run from $1$ to $3$. The signature of the spacetime metric is $+2$.

\section{Static Observers and Adapted Frames}

Observational data for AGNs refer to the rest frame of the ``central engine". We consider the family of static observers in the spacetime exterior to the gravitational source to represent this rest frame \emph{locally}. More specifically, the purpose of this section is to study static observers in the exterior  Kerr spacetime  with non-rotating spatial  frames along their world lines. 

\subsection{Static Observers}

Any test family of observers is characterized by a unit timelike 4-velocity field $u^\mu$ whose integral curves are the world lines of the observers. The static observers in the exterior Kerr spacetime follow the integral curves of the (stationary) Killing vector field $\partial_t$, namely, 
\begin{equation}\label{S1}
u^\mu=\frac{dx^\mu}{d\tau}\,, \qquad u=\left(\frac{\Sigma}{\Sigma - 2\,M\,r}\right)^{1/2}\,  \partial_t\,.
\end{equation}
Only positive square roots are considered throughout. Moreover, $\tau$ is the proper time along the path,
\begin{equation}\label{S2}
\tau =\left(\frac{\Sigma- 2\,M\,r}{\Sigma}\right)^{1/2}\,t\,,
\end{equation}
where we have assumed that $\tau=0$ at $t=0$. The natural orthonormal tetrad frame $e^{\mu}{}_{\hat \alpha}$ of these observers is given by~\cite{Bini:2016xqg}
\begin{eqnarray}\label{S3}
\nonumber e_{\hat 0}= u\,, \qquad e_{\hat 1}= \left(\frac{\Delta}{\Sigma}\right)^{1/2}\,\partial_r\,, \qquad  e_{\hat 2}= \left(\frac{1}{\Sigma}\right)^{1/2}\,\partial_\theta\,,\\
e_{\hat 3} =\frac{- 2\,M\,a\,r\,\sin \theta}{\left[\Delta\,\Sigma\,(\Sigma- 2\,M\,r)\right]^{1/2}}\,  \partial_t+ 
\left(\frac{\Sigma- 2\,M\,r}{\Delta\,\Sigma }\right)^{1/2}\,\frac{1}{\sin \theta}\,\partial_\phi\,,
\end{eqnarray}
where the tetrad axes are primarily along the Boyer-Lindquist coordinate directions.  
The static observers form a congruence of accelerated, non-expanding and locally rotating world lines. The lack of expansion of the congruence is due to the alignment of its 4-velocity vector field with the timelike Killing direction. 

It is straightforward to check that the 4-acceleration of the static observers is given by 
\begin{equation}\label{S4}
\mathcal{A}^\mu=\frac{De^{\mu}{}_{\hat 0}}{d\tau} = \Gamma^{\mu}_{\alpha \beta}\,e^{\alpha}{}_{\hat 0}\,e^{\beta}{}_{\hat 0}\,,
\end{equation}
where $\Gamma^{\mu}_{\alpha \beta}$ are the connection coefficients for the Kerr spacetime in Boyer-Lindquist coordinates. Thus, 
\begin{eqnarray}\label{S5}
\mathcal{A}&=&\frac{M\sqrt{\Delta}(r^2-a^2\cos^2\theta)}{\Sigma^{3/2}(\Delta -a^2 \sin^2\theta)}\,e_{\hat 1}\nonumber\\
&& -
\frac{2Mra^2 \sin \theta \cos \theta}{\Sigma^{3/2}(\Delta -a^2 \sin^2\theta)}\,e_{\hat 2}\,.
\end{eqnarray}
This acceleration, due to forces that are not gravitational in origin,  is necessary to keep the reference observer static and prevent it from falling into the source. 
\subsection{Fermi-Walker Frame}

The next step involves the establishment of a Fermi-Walker transported spatial frame along the world line of each static observer. Let $S^\mu$ be a vector that is Fermi-Walker transported along  $e^{\mu}{}_{\hat 0}$; then,  
\begin{equation}\label{F1}
\frac{dS^\mu}{d\tau}+\Gamma^{\mu}_{\alpha \beta}\, e^{\alpha}{}_{\hat 0}\,S^{\beta}= (\mathcal{A}\cdot S)\,e^{\mu}{}_{\hat 0}-(e_{\hat 0} \cdot S)\,\mathcal{A}^\mu\,.
\end{equation}
Let us assume that $S$ is orthogonal to the world line, i.e. $e_{\hat 0} \cdot S=0$. Moreover, it proves useful to express $S$ in terms of the natural spatial frame $e^{\mu}{}_{\hat a}$; hence,
\begin{equation}\label{F2}
S^\mu = s^{\hat a}\,e^{\mu}{}_{\hat a}\,.
\end{equation}
We now write Eq.~\eqref{F1} in terms of $s^{\hat a}$ and obtain, after some algebra,  
\begin{equation}\label{F3}
\frac{ds^{\hat 1}}{dt} = \frac{M\,a\,(r^2-a^2 \cos^2 \theta)\,\sin \theta}{\Sigma^2\,(\Sigma- 2\,M\,r)^{1/2}}\,s^{\hat 3}\,,
\end{equation}
\begin{equation}\label{F4}
\frac{ds^{\hat 2}}{dt} = - \frac{2\,M\,a\,r\,\sqrt{\Delta}\,\cos \theta}{\Sigma^2\,(\Sigma- 2\,M\,r)^{1/2}}\,s^{\hat 3}\,,
\end{equation}
\begin{equation}\label{F5}
\frac{ds^{\hat 3}}{dt} = \frac{M\,a}{\Sigma^2\,(\Sigma- 2\,M\,r)^{1/2}}\,[-(r^2-a^2 \cos^2 \theta)\,\sin \theta \,s^{\hat 1} + 2\,r \sqrt{\Delta}\,\cos \theta \, s^{\hat 2}]\,.
\end{equation}
It proves convenient to define an angle $\alpha$ and a precession frequency $\beta>0$ such that
\begin{eqnarray}\label{F6}
\cos\alpha &=& \frac{2\,r\,\sqrt{\Delta} \,\cos\theta}{[4\,r^2 \Delta\cos^2\theta + (r^2-a^2\cos^2\theta)^2\,\sin^2\theta]^{1/2}}\,,\nonumber\\
\sin\alpha &=& \frac{(r^2-a^2\cos^2\theta)\,\sin \theta}{[4\,r^2 \Delta\cos^2\theta + (r^2-a^2\cos^2\theta)^2\,\sin^2\theta]^{1/2}}\,
\end{eqnarray}
and 
\begin{equation}\label{F7}
\beta = \frac{Ma}{\Sigma^2}\left[\frac{4\,r^2 \Delta\cos^2\theta + (r^2-a^2\cos^2\theta)^2\,\sin^2\theta}{\Sigma-2\,M\,r}\right]^{1/2}\,.
\end{equation}
We note that $\alpha$ has the same range as the polar angle $\theta$; that is, 
$\alpha:0\to \pi$ when $\theta: 0 \to \pi$.
Equations~\eqref{F3}--\eqref{F5} can now be expressed as 
\begin{equation}\label{F8}
\frac{ds^{\hat i}}{dt} =\epsilon^{\hat i \hat j \hat k}\,\Omega_{\hat j}\,s_{\hat k}\,, 
\end{equation}
where
\begin{equation}\label{F9}
\Omega^\mu = \Omega^{\hat a}\, e^{\mu}{}_{\hat a} = \beta (\cos \alpha \, e^{\mu}{}_{\hat 1} + \sin \alpha \, e^{\mu}{}_{\hat 2})
\end{equation}
is the precession vector. 
On the basis of these results, we can straightforwardly construct a locally non-rotating spatial frame $\lambda^{\mu}{}_{\hat a}$ along $e^{\mu}{}_{\hat 0}$. For instance, let $\lambda^{\mu}{}_{\hat 1}$ be the unit vector along $\Omega^\mu$; then, $\lambda^{\mu}{}_{\hat 2}$ and $\lambda^{\mu}{}_{\hat 3}$ are unit vectors in the plane orthogonal to $\lambda^{\mu}{}_{\hat 1}$ and precess with frequency $\beta$ about $\lambda^{\mu}{}_{\hat 1}$. More explicitly, consider first the orthonormal spatial frame $E^{\mu}{}_{\hat a}$,
\begin{eqnarray}\label{F10}
E_{\hat1} &=& \cos \alpha\, e_{\hat 1}+\sin \alpha\, e_{\hat 2}\,, \nonumber\\
E_{\hat 2} &=&-\sin \alpha\, e_{\hat 1}+\cos \alpha\, e_{\hat 2}\,, \nonumber\\
E_{\hat 3} &=& e_{\hat 3}\,.
\end{eqnarray}
Then the  Fermi-Walker triad $\lambda^{\mu}{}_{\hat a}$ is obtained from $E^{\mu}{}_{\hat a}$  by a simple rotation about $E^{\mu}{}_{\hat 1}$ with an angle of $\beta t$,
\begin{eqnarray}
\label{F11}
\lambda_{\hat 1} &=& E_{\hat 1}\,,  \nonumber\\
\lambda_{\hat 2} &=&\cos (\beta t) \, E_{\hat 2}+\sin (\beta t) \, E_{\hat 3}\,, \nonumber\\
\lambda_{\hat 3} &=& -\sin (\beta t) \, E_{\hat 2}+\cos (\beta t) \, E_{\hat 3}\,.
\end{eqnarray}
This completes the construction of the non-rotating tetrad frame  $\lambda^{\mu}{}_{\hat \alpha}$, where $\lambda^{\mu}{}_{\hat 0}=e^{\mu}{}_{\hat 0}$ along the world line of a static observer in the exterior Kerr spacetime. 

Appendix A is devoted to the construction of a geodesic (Fermi) coordinate system along the locally non-rotating tetrad frame $\lambda^{\mu}{}_{\hat \alpha}$ in the neighborhood of the corresponding static observer. Henceforth, we consider the motion of free test particles in this Fermi normal coordinate system. 

\section{Geodesic Motion in Fermi Coordinates}

Using the Fermi-Walker transported frame $\lambda^{\mu}{}_{\hat \alpha}(\tau)$ along the world line of the reference observer, a Fermi normal coordinate system $X^\mu = (T, \mathbf{X})$ can be established in its neighborhood---see Appendix A. This system makes it possible to provide an invariant description of the motion of the test particles relative to the fiducial observer that occupies the origin of the spatial Fermi coordinates and locally represents the rest frame of the gravitational source. 

For simplicity we neglect plasma effects in this work; therefore, a test particle follows a geodesic path that can be expressed in Fermi coordinates as
\beq\label{G1}
\frac{d^2 X^\mu}{ds^2}+\tilde{\Gamma}^\mu{}_{\alpha\beta}\,\frac{dX^\alpha}{ds}\,\frac{dX^\beta}{ds}=0\,.
\eeq
The free test particle has 4-velocity $U^\mu$,
\beq\label{G2}
U^\mu := \frac{dX^\mu}{ds} = \Gamma\,(1, \mathbf{V})\,,
\eeq
where $U^\mu\,U_\mu = -1$ implies that the Lorentz factor  $\Gamma$ is given by
\beq\label{G3}
\Gamma = \frac{1}{(-\tilde{g}_{00}-2\,\tilde{g}_{0i}\,V^i -\tilde{g}_{ij}\,V^i\,V^j)^{1/2}}\,. 
\eeq
In the immediate neighborhood of the reference observer, the space is Euclidean and Fermi velocity $\mathbf{V}$ of the test particle must satisfy the condition that $|\mathbf{V}| \le 1$ at $\mathbf{X}=0$. 

Separating Eq.~\eqref{G1} into its temporal and spatial  components, it is straightforward to obtain   the  reduced geodesic equation~\cite{CM1}
\beq\label{G4}
\frac{d^2 X^i}{dT^2}+\left(\tilde{\Gamma}^i{}_{\alpha\beta}-\tilde{\Gamma}^0{}_{\alpha\beta}V^i \right) \frac{dX^\alpha}{dT}\frac{dX^\beta}{dT}=0\,.
\eeq
A detailed discussion of motion in the Fermi coordinate system is contained in Refs.~\cite{CM1, CM2, CM3}. For instance, as described in Ref.~\cite{CM1}, Eq.~\eqref{G4} can also be obtained via the isoenergetic reduction procedure from Eq.~\eqref{G1}.

Inserting the expressions for the Christoffel symbols in Fermi coordinates $\tilde{\Gamma}^{\mu}{}_{\alpha\beta}$ given in Appendix A, we find that Eq.~\eqref{G4} can be written as
\begin{eqnarray}\label{G5}
&&\frac{d^2X^i}{dT^2}+A_i+(\tilde{R}_{0i0l}+A_iA_l)X^l-\frac{dA_l}{dT}X^l V^i\nonumber\\
&& \qquad-2\,\tilde{R}_{0lij}V^jX^l-2\,[A_j+(\tilde{R}_{0j0l}-A_jA_l)\,X^l]V^iV^j\nonumber\\
&& \qquad -\frac23 \left(\tilde{R}_{ijkl}+\tilde{R}_{0jkl}V^i \right) X^lV^jV^k + O(|\mathbf{X}|^2)=0\,.
\end{eqnarray} 
Moreover, since the motion is timelike we have
\begin{equation}\label{G6}
\frac{1}{\Gamma^2} =  (1+A_i\,X^i)^2 -V^2 + \tilde{R}_{0i0j}\,X^i\,X^j  +\frac{4}{3}\, \tilde{R}_{0jik}V^i  X^j\,X^k+ \frac13\,\tilde{R}_{ikjl}\,V^i\,V^j\,X^k\,X^l +O(|\mathbf{X}|^3)> 0\,,
\end{equation}
where $V^2= \delta_{ij}\,V^i\,V^j$. It is useful to define $V_i := \delta_{ij}\,V^j$, so that $V^2=V_i\,V^i$.

\section{Motion Along the Jet}

Let us now specialize these general results to motion along the jet direction. Imagine a reference observer at rest along the rotation axis of a Kerr source and focus on the motion of free test particles relative to the reference observer along the positive jet direction. To this end, we must limit Eqs.~\eqref{G5} and~\eqref{G6} to motion along one spatial direction, so that the only non-zero components of   $\mathbf{X}$ and $\mathbf{V}$ are $X^1$ and $V_1$, respectively, since the Boyer-Lindquist polar angle vanishes along the positive jet direction ($\theta=0$). Furthermore, as explained in detail in Appendix B, it turns out that
\begin{eqnarray}\label{J1}
A_1&=& \frac{M(r^2-a^2)}{(r^2+a^2)^{3/2}\Delta^{1/2}}\,, \nonumber\\
\tilde{R}_{0101}&=& -\frac{2Mr (r^2-3a^2)}{(r^2+a^2)^3}\,,
\end{eqnarray} 
where $r$ is the fixed Boyer-Lindquist radial coordinate that here represents the location of the reference observer on the rotation axis of the central source. The acceleration and curvature terms in Eq.~\eqref{J1} only depend upon the square of the magnitude of the specific angular momentum of the source and not on its directionality. Since $A_1$ is independent of time $T$, Eqs.~\eqref{G5} and~\eqref{G6} reduce to
\beq 
\label{J2}
\frac{d^2X^1}{dT^2}+(A_1+\tilde{R}_{0101}\,X^1) (1-2 V_1^2)+A_1^2\, X^1\,(1+2 V_1^2)+O(|\mathbf{X}|^2)=0\,
\eeq
and 
\beq\label{J3}
\Gamma^{-2}=(1+A_1\,X^1)^2-V_1^2+\tilde{R}_{0101}\,(X^1)^2 +O(|\mathbf{X}|^3)> 0\,.
\eeq

Let us briefly digress here and mention that in the absence of acceleration $A_1$, Eq.~\eqref{J2} reduces to the generalized Jacobi equation that has been the subject of a number of previous investigations---see~\cite{CM1, CM2, CM3} and the references cited therein. Those studies revealed the existence of an \emph{attractor} in the system, namely, motion at the terminal speed $c/\sqrt{2} \approx 0.7\,c$~\cite{CM1, CM2, CM3}. However, it turns out that the nature of the motion changes completely in the presence of $A_1$. The main purpose of the present work is to study jet motion when the fiducial observer is accelerated.

An important feature of Eqs.~\eqref{J2} and~\eqref{J3} must be mentioned here: They remain invariant under the transformation
\beq\label{J3a}
X^1 \mapsto -X^1\,, \qquad A_1 \mapsto -A_1\,, \qquad \tilde{R}_{0101} \mapsto \tilde{R}_{0101}\,.
\eeq
As discussed in Appendix B, a consequence of Eq.~\eqref{J3a} is that the tidal dynamics of the jet outflow in the northern hemisphere of the source is precisely the same as for the jet moving in the southern hemisphere. Henceforth, we will concentrate on the jet in the northern hemisphere of the gravitational source. 

To proceed, we define constant positive dimensionless parameters $\Phi$ and $\eta$ such that 
\beq\label{J4}
\Phi=\frac{GM}{c^2r}\,,\qquad \eta=\frac{J}{Mc\,r}=\frac{a}{r}\,,
\eeq
where $J = M c \,a$ is the angular momentum of the gravitational source. Next, we define dimensionless quantities $p$ and $q$, 
so that
\beq\label{J5}
A_1 :=\frac{GM}{c^2 r^2}\,p\,,\qquad
\tilde{R}_{0101} := -\frac{GM}{c^2 r^3}\,q\,.
\eeq
It follows from Eq.~\eqref{J1} that 
\begin{eqnarray}\label{J6}
p&=& \frac{1-\eta^2}{(1+\eta^2)^{3/2}\,(1-2\,\Phi+\eta^2)^{1/2}}\,,\nonumber\\
q&=& 2\,\frac{1-3\eta^2}{(1+\eta^2)^3}\,.
\end{eqnarray}
To express Eqs.~\eqref{G5} and~\eqref{G6} in terms of dimensionless quantities, let
\beq\label{J7}
X^1 := r\, z\,, \qquad 
T :=\frac{r}{\sqrt{\Phi}}\,\zeta\,.
\eeq
Then, 
\beq\label{J8}
V_1=\frac{dX^1}{dT}= \sqrt{\Phi}\,\,\frac{dz}{d\zeta}
\eeq
and we find
\begin{eqnarray}
\label{J9}
&& \frac{d^2z}{d\zeta^2}+(p-qz)\left[ 1-2\Phi \left(\frac{dz}{d\zeta}\right)^2 \right]\nonumber\\
&& \qquad +\Phi p^2 z  \left[ 1+2\Phi \left(\frac{dz}{d\zeta}\right)^2 \right] + O(z^2)=0\,,
\end{eqnarray}
\beq\label{J10}
\Gamma^{-2}=(1+\Phi p z)^2- \Phi\, \left(\frac{dz}{d\zeta}\right)^2 - \Phi q z^2 + O(z^3)> 0\,.
\eeq

In the neighborhood of the reference observer, the radius of curvature of spacetime is 
\begin{equation}\label{J11}
\mathcal{R} \sim \frac{r}{\sqrt{\Phi}}\,;
\end{equation}
therefore, we expect that Eq.~\eqref{J9} describes the motion of free test particles properly only for 
\begin{equation}\label{J12}
|z| <  \frac{1}{\sqrt{\Phi}}\,.
\end{equation}
Moreover, it is very important to recognize that only the \emph{tidal effects} of the central source are represented in the metric of the Fermi coordinate system. The source of the gravitational field itself does not appear in this analysis, since our treatment is restricted to the exterior of the source. 

We now turn to the analysis of jet motion in accordance with Eq.~\eqref{J9}.

\section{Phase Plane Analysis}\label{sec:pp}

%%%%%%%%%%%%%%%
%%%%%%%%%%%%%%
\begin{figure}
\begin{center}
\includegraphics[scale=1]{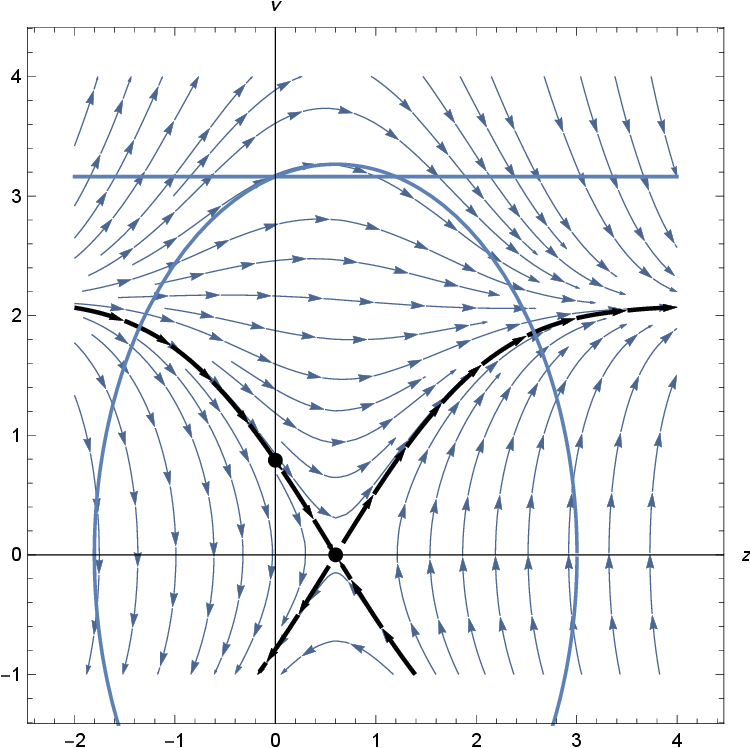}  
\caption{\label{fig.1}  The phase portrait of ODE~\eqref{P1} is depicted  for the case $\Phi=1/10$ and $\eta=1/20$. The vertical coordinate axis corresponds to  $v:=dz/d\zeta$. Also shown are the  ellipse~\eqref{P6} corresponding to $\Gamma=\infty$,  the position of the hyperbolic saddle rest point ($z=z_0$),  the critical value of $v_0$ along the vertical axis ($v_0=v_{ta}$) and the line $v=1/\sqrt{\Phi}$.  The initial outflow velocity $v_0$ at $z=0$ is such that $0 < v_0 < 1/\sqrt{\Phi}$. The phase trajectory through the critical point is shown with a  thick curve.  It is part of the stable manifold of the saddle point. For $v_0>0$, all solutions starting below the critical value  reach a maximum $z$ at a point where $v=0$. All solutions starting above the critical value and below the horizontal line $v=1/\sqrt{\Phi}$ eventually meet the ellipse corresponding to $\Gamma=\infty$. The critical value for tidal acceleration in the outflow, $v_{ta} \approx 0.79$, corresponds in this case to a Fermi speed of $V_{ta} \approx 0.25\, c$.}
\end{center}
\end{figure}
%%%%%%%%%%%%%%%%%%
%%%%%%%%%%%%%%

It is revealing to consider the ordinary differential Eq.~\eqref{J9} to $O(z)$  in the phase plane, where it is viewed as the first-order system~\cite{CH} 
\begin{equation}\label{P1}
\frac{dz}{d\zeta} = v\,, \qquad \frac{dv}{d\zeta} = (-p+qz) (1-2\Phi v^2) - \Phi p^2 z (1+2\Phi v^2)\,. 
\end{equation}
This system has a rest point at $z=z_0$, where
\begin{equation}\label{P2}
z_0 = \frac{p}{q-\Phi\,p^2}\,. 
\end{equation}
That is, system~\eqref{P1} has a simple solution given by $z=z_0$ and $v=0$. In most astrophysical contexts, we expect that 
\begin{equation}\label{P3}
\Phi \ll1\,, \qquad   \eta \ll 1\,,
\end{equation}
so that $p$, $q/2$ and $2\,z_0$ are all nearly equal to unity.  This case is considered in  the remainder of this section. 

We are interested in trajectories of free test particles in the jet outflow that reach the reference observer at $z=0$ and have at this point an initial speed $v_0$,
\begin{equation}\label{P4}
v_0 :=v (z=0)\,,
\end{equation} 
such that 
\begin{equation}\label{P5}
0 < v_0 < \frac{1}{\sqrt{\Phi}}\,, 
\end{equation}
since the (Fermi) speed of the particle relative to the reference observer at $z=0$, $V_0=\sqrt{\Phi}\,v_0$,  must be less than the speed of light. 
In particular, we would like to know if such a test particle is tidally accelerated. If the test particle is accelerated to the speed of light, then  in that limit it approaches  $\Gamma = \infty$.  If follows from  
Eq.~\eqref{J10} that to $O(z^2)$, the $\Gamma = \infty$ locus is an ellipse given by 
\begin{equation}\label{P6}
 (1+\Phi p z)^2- \Phi v^2 - \Phi q z^2 =0\,.
\end{equation}
This equation can be written as  
\begin{equation}\label{P7}
\frac{v^2}{\mathbb{A}^2} +\frac{(z-z_0)^2}{\mathbb{B}^2}=1\,,
\end{equation}
where the semimajor axis $\mathbb{A}$ and the semiminor axis $\mathbb{B}$ of the ellipse are given by
\begin{equation}\label{P8mm}
\mathbb{A} =\left[\frac{q}{(q-\Phi\,p^2)\,\Phi}\right]^{1/2}\,, \qquad \mathbb{B}=\left[\frac{q}{(q-\Phi\,p^2)^2\,\Phi}\right]^{1/2}\,.
\end{equation}

A typical example is shown in Figure 1.  All features of the phase portrait can be determined by elementary means because  system~\eqref{P1}  has a first integral, which is obtained via a simple observation: the ordinary differential equation (ODE)
\begin{equation}\label{P7b}
\frac{1}{2} \frac{d}{dz} v^2= v \frac{dv}{dz}= (-p+qz) (1-2\Phi v^2) - \Phi p^2 z (1+2\Phi v^2)
\end{equation}
is linear in $v^2$, see Appendix C.  For solutions starting with $z=0$ and $v_0>0$,  there is a critical value of  $v_0$, $v_0=v_{ta}$,  below which they reach a maximum $z < z_0$ and thereafter the direction of $v$ is reversed and they fall back toward the Kerr source.  Solutions of this model starting above this critical value $v_{ta}$ and below the value of $v_0$ where the model is physically relevant ($v_0=1/\sqrt{\Phi}$) evolve in forward time to a point where $\Gamma=\infty$ in this approximation; that is, these solutions reach the ellipse~\eqref{P6}. 

In Figure 1, the hyperbolic saddle resides at the rest point with  coordinates $(z_0,0)$. The solution starting at the critical point with coordinates $(0,v_{ta})$ lies on the stable manifold of the rest point.   The critical velocity 
$v_{ta}$ is given by
\begin{equation}\label{P8}
v_{ta}^2=2 \int_0^{z_0} [ p- (q-\Phi p^2)z]\, e^{-4 \Phi p z+2 \Phi( q+\Phi p^2) z^2}\,dz\,,
\end{equation}
which can be obtained from Eq.~\eqref{C9} of Appendix C by setting $v(z_0) =0$ and defining the corresponding $v_0$ to be the critical velocity for tidal acceleration $v_{ta}$.  It is possible to develop a series expression for $v_{ta}$, namely,  
\begin{equation}\label{P8a}
v_{ta} = \frac{1}{\sqrt{2}} \,(1+\Phi) + O(\Phi^{3/2}) + O(\eta^2)\,.
\end{equation}

We recall from Eq.~\eqref{J8} that the actual (Fermi) velocity is given by
\begin{equation}\label{P8b}
V = \sqrt{\Phi}\, v\,;
\end{equation}
therefore, the actual critical velocity for tidal acceleration is 
\begin{equation}\label{P8c}
V_{ta} = \sqrt{\frac{\Phi}{2}} \,(1+\Phi) + \cdots\,.
\end{equation}
Thus, to lowest order in the small quantities $\Phi \ll 1$ and $\eta \ll 1$, the critical speed for tidal acceleration is one half of the Newtonian escape velocity $\sqrt{2\,\Phi}\,c$ at the location of the reference observer.

\subsection{$v_{ta} < v_0 < \frac{1}{\sqrt{\Phi}}$}

If the speed of the free test particle that reaches the location of the reference observer is above the critical speed $v_{ta}$, then the particle is tidally accelerated to almost the speed of light, see Figure 2.  

%%%%%%%%%%%%%%%%%%%
%%%%%%%%%%%%%%%%%%%%
\begin{figure}
\begin{center}
\includegraphics[scale=0.5]{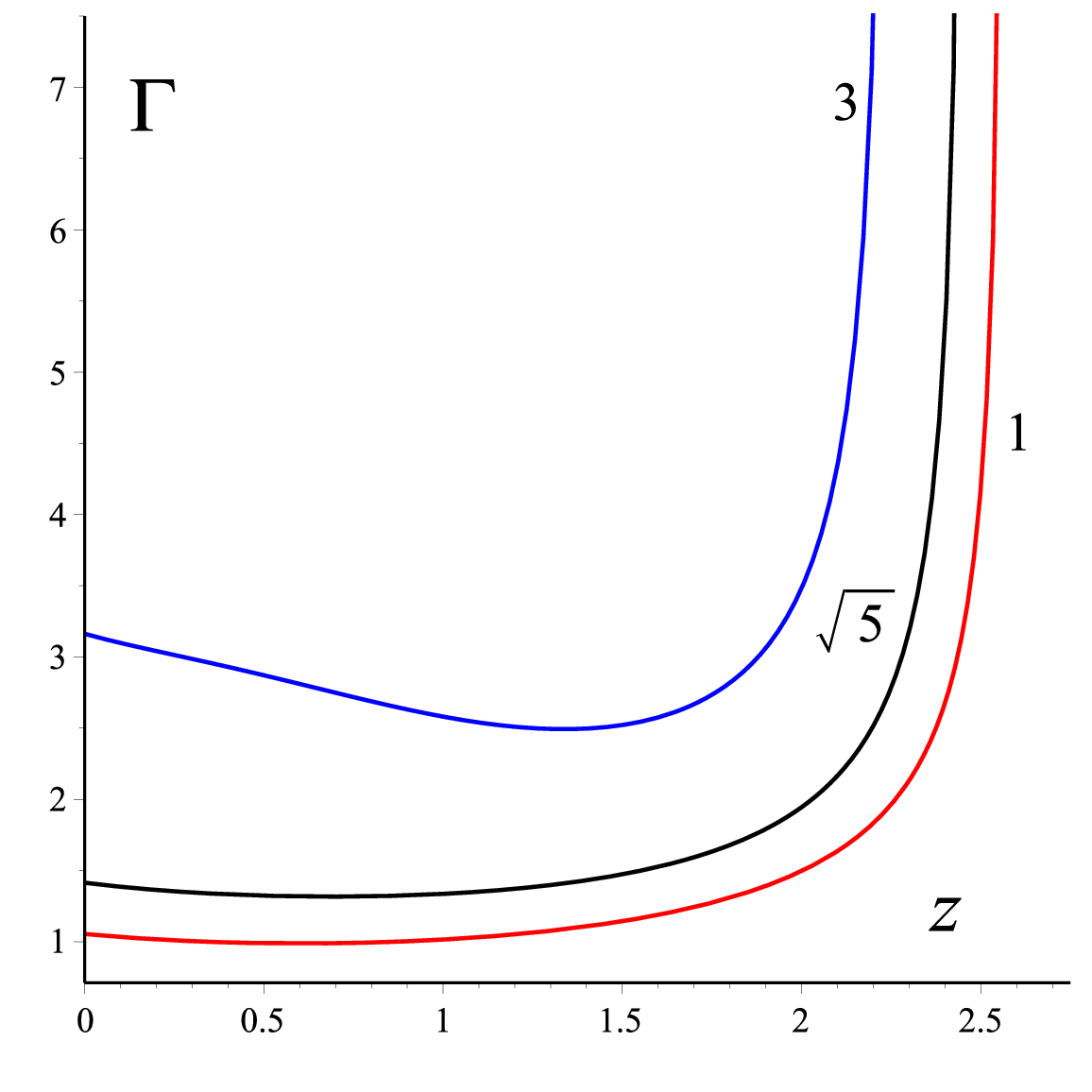} 
\caption{\label{fig.2} The behavior of the Lorentz factor $\Gamma$ is shown versus $z$ by using the numerical integration of system~\eqref{P1} for $\Phi=1/10$ and $\eta=1/20$. The initial conditions are $z=0$ 
and $v_0=1,\sqrt{5},3$.
}
\end{center}
\end{figure}
%%%%%%%%%%%%%%%%%%%%
%%%%%%%%%%%%%%%%%%%%

It is important to note that the axes of the $\Gamma=\infty$ ellipse lie at the boundary of the cylindrical region where Fermi coordinates are admissible. That is, it follows from Eqs.~\eqref{J11} and~\eqref{P8mm} that 
\begin{equation}\label{P9}
 r\,\mathbb{A} \sim r\,\mathbb{B}  \sim \frac{r}{\sqrt{\Phi}}\,\sim \mathcal{R}\,,
\end{equation}
so that as the particle is accelerated to $\Gamma \gg 1$, the higher-order tidal terms that have been thus far neglected (as well as plasma effects) enter the analysis and mitigate the $\Gamma=\infty$ singularity, which the test particle never in fact reaches. 

\subsection{$0 < v_0 < v_{ta}$}

If the speed of the free test particle in the jet outflow is less than the critical speed for tidal acceleration, the particle moves past the reference observer and reaches a maximum height of $z<z_0$. It then falls back toward the gravitational source. As it falls, it experiences tidal acceleration away from the reference observer and toward the source. According to Figure 1, the particle would eventually reach the 
$\Gamma=\infty$ ellipse; however, this conclusion is in any case illusory, especially since the tidal approach loses its significance when the particle gets very close to the source. 

\section{Special Cases}

The mathematical model under consideration in this paper is applicable for a wider range of parameter values than those discussed in Section~\ref{sec:pp}. For the sake of completeness, it is interesting to investigate certain special cases as well. For instance, 
if the gravitational source is an extreme Kerr black hole ($M = a$),  then $\eta= \Phi$ and the fiducial  static observer can in principle occupy positions along the rotation axis all the way down to just outside $r=M$. Therefore,  we must have $\Phi=\eta <1$ and  
\begin{equation}\label{P10}
 p=\frac{1+\Phi}{(1+\Phi^2)^{3/2}}\,, \qquad  q=2\,\frac{1-3 \Phi^2}{(1+\Phi^2)^3}\,.
\end{equation}
In this case, $z_0$ increases from $1/2$ at $\Phi=0$ and approaches $\infty$ as $\Phi \to \Phi_c \approx 0.43$. For $\Phi_c < \Phi <1$, $z_0$ is negative. 

Another possibility involves a rapidly rotating Kerr source ($a \gg M$) with, say,  $\Phi \ll1$, but $\eta \sim 1$. We note that $q=0$ for $\eta=1/\sqrt{3}$ and $q<0$ for $\eta>1/\sqrt{3}$.

%%%%%%%%%%%%%%%%%
%%%%%%%%%%%%%%%%%%
\begin{figure}
\begin{center}
\includegraphics[scale=1]{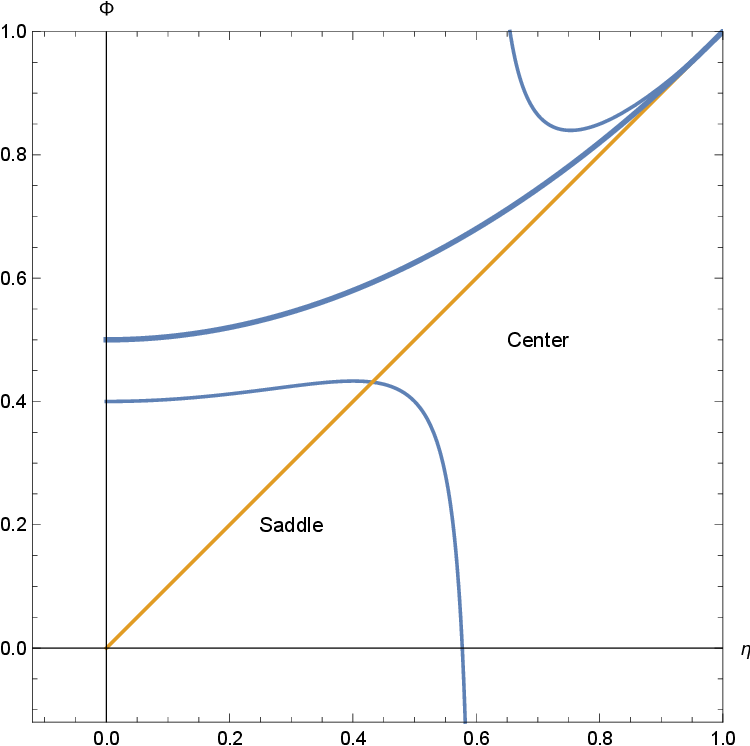}  
\caption{\label{fig.bd}  The bifurcation diagram for the phase portrait of system~\eqref{P1}  with respect to the parameters $\Phi$ and $\eta$ is shown. In this $(\eta, \Phi)$ plane, the thick curve, which corresponds to the locus of the event horizon $1-2\,\Phi + \eta^2=0$,  is the upper boundary of the relevant parameter region bounded also by the coordinate axes and the line $\eta=1$. The line $\Phi=\eta$ is depicted as well as the bifurcation curve corresponding to $q-\Phi\, p^2=0$. Below the lower branch of the latter curve, the phase portrait has a saddle-type rest point; above the curve it has a center. The upper branch is also depicted, but lies outside the domain of physical interest.}
\end{center}
\end{figure}
%%%%%%%%%%%%%%%
%%%%%%%%%%%%%%%

\begin{figure}
\begin{center}
\includegraphics[scale=1]{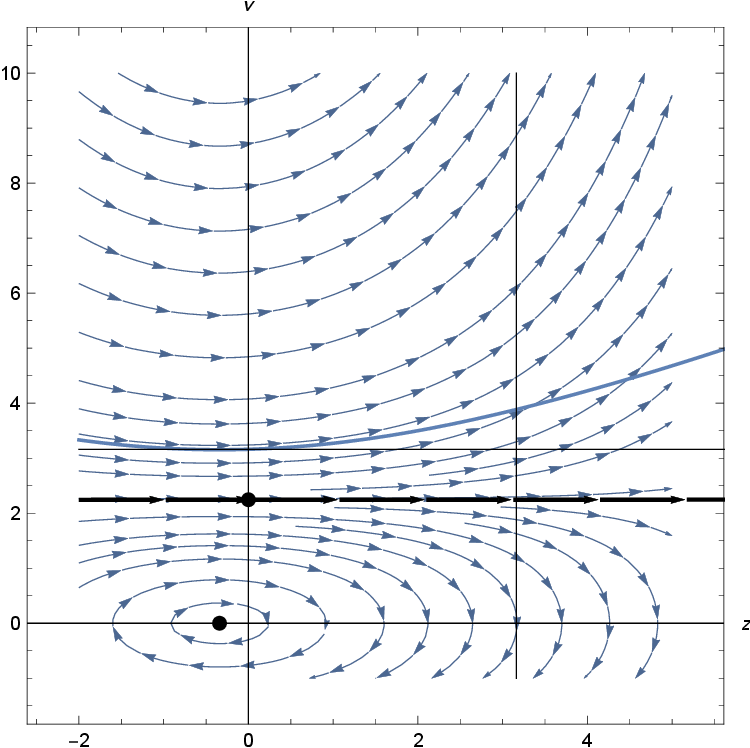}  
\caption{\label{fig.center}  The phase portrait  for system~\eqref{P1} where $q-\Phi\, p^2<0$ is depicted for the rapidly rotating case of $\Phi=1/10$ and $\eta=8/10$. The upper branch of the hyperbola corresponding to $\Gamma=\infty$ is also shown. Also,  the rest point, the critical point on the vertical axis, and the critical curve are depicted. The line $v =1/\sqrt{\Phi}$ shows the limiting value of $v_0$, while the line $z=1/\sqrt{\Phi}$ indicates the outer boundary of the admissibility of Fermi coordinates. The critical value for outflow tidal acceleration, $v_{ta} \approx 2.245$, corresponds in this case to a (Fermi) speed of $V_{ta} \approx 0.71\,c$. We note that in this case the outflow tidal acceleration is in fact insignificant in comparison to the case presented in Figure 1.}
\end{center}
\end{figure}

When we take such special cases into account, the phase portrait of  system~\eqref{P1} does not remain qualitatively the same  as the parameters $\Phi$ and $\eta$ vary beyond $\Phi\ll 1$ and $\eta\ll 1$. Indeed its  signature feature, the nature of the rest point, goes through a bifurcation as the value of $q-\Phi p^2$ passes through zero.    The corresponding bifurcation diagram is presented in  Figure~\ref{fig.bd}. This issue is discussed in detail in Appendix C; indeed, in this transition a certain effective potential energy function defined in Appendix C changes from a potential \emph{barrier} to a potential \emph{well}. In the region marked ``Saddle'' in Figure 3,  the phase portrait of the barrier is qualitatively the same as that in Figure~\ref{fig.1}.   In the region marked ``Center'',   the phase portrait of the well is qualitatively the same as in Figure~\ref{fig.center}. 

%%%%%%%%%%%%%%%%
%%%%%%%%%%%%%%%%
%%%%%%%%%%%%%
%%%%%%%%%%%%%%

In case the rest point is a center, it resides on the negative $z$ axis or at the origin.  The conic section~\eqref{P6} corresponding to an infinite Lorentz gamma factor is a hyperbola in this case. Again the basic scenario for solutions starting at $z=0$ and positive $v_0$ is the same: For small $v_0$ below some critical value, the solutions---which are on periodic orbits 
of the first-order system~\eqref{P1}---reach a maximum value of $z$ at a point where $v=0$ and the motion continues with the sign of $v$ reversed. The evolving Lorentz gamma factor for solutions starting with $v_0$ above the critical value reaches infinity in some finite time  corresponding to the moment the phase plane trajectory reaches the upper branch of the hyperbola.  The critical value on the vertical axis is given by formula~\eqref{P8} with $z_0$ replaced by $\infty$. The latter result can be proved  using the first integral derived in Appendix C---in fact, the situation is illustrated in Figure 6. 
The structure of the phase portrait in this case can be understood based on the results presented in Appendix C. In particular,  the annulus of periodic solutions surrounding the center has infinite extent in the positive direction and finite extent in the negative direction.  Because the annulus has infinite extent in the positive direction, the critical value $v_{ta}$ is indeed given in this case by Eq.~\eqref{P8} with $z_0$ replaced by $\infty$.

Let us return briefly to the two physically important cases involving the extreme Kerr black hole ( $\Phi=\eta <1$)  and the rapidly rotating Kerr source ($\Phi \ll1$, but $\eta \sim 1$).  In the former case, the nature of the phase portrait is determined by position on the line $\Phi=\eta$ according to the bifurcation diagram in Figure~\ref{fig.bd}. For small $\eta$, the saddle case is predicted;  for $\eta$ above the bifurcation curve (at $\approx 0.43$) the center case occurs.  In the latter case, the rest point is a center that resides near the origin as in Figure~\ref{fig.center} and the critical speed for tidal acceleration is approximately $v_{ta}=2.245$. 

Suppose we fix the value of $\Phi$ such that $\Phi \ll1$ and let $\eta$ vary from $0$ to $1$. The question is what happens to the value of $v_{ta}$? Numerical studies show that well within the saddle and center regimes in Figure 3, $v_{ta}$ is nearly constant, but experiences a significant jump in going from the saddle regime to the center regime. For instance, for $\Phi=0.1$, $v_{ta}$ jumps from roughly around 0.8 to roughly around 2 when $q-\Phi\,p^2$ goes from positive to negative. This seems to agree with observational data~\cite{Fender:2014sia} that there is in effect  no significant correlation between jet activity and the rotation of the black hole, since for $\eta$ from $0$ to $\Phi$, we are well  within the saddle regime and $v_{ta}$ is essentially constant.

\section{DISCUSSION}

Imagine a Kerr source with mass $M$, angular momentum $J=Ma\,c$ and bipolar outflows along its rotation axis. We are interested in the tidal acceleration of jet particles relative to the central Kerr source and, for the sake of simplicity, we concentrate on particles moving out along the positive direction of the rotation axis. We pick a position  $r \gg GM/c^2$ on the rotation axis away from the Kerr source, where $r$ is the radial Boyer-Lindquist coordinate. This is the location of a hypothetical fiducial observer that is at rest in the exterior Kerr spacetime and provides us with a \emph{local} representation of the rest frame of the source. We would like to know if the outflow particles that reach this point are tidally accelerated as they move forward.  All electromagnetic forces are neglected in our analysis for the sake of simplicity. Regarding the rotation of the Kerr source, we consider two cases: (i) a moderately rotating source with $0 \le a \ll r$, which includes the extreme Kerr black hole and beyond and (ii) a very rapidly rotating Kerr source with specific angular momentum $a\gg GM/c^2$ including, say, $a \sim r$.  
Let $V_0$ be the (Fermi) velocity of a particle that reaches the point under consideration. In the first case involving a moderately rotating Kerr source, we find that if $V_0$ is above a certain threshold, roughly corresponding to one half of the \emph{Newtonian escape velocity $V_N$}
\begin{equation}\label{P11}
V_N := \left(\frac{2GM}{r}\right)^{1/2}\,,
\end{equation}
then the particle can be tidally accelerated to almost the speed of light. If $V_0$ is below this threshold, then the particle reaches a maximum height and falls back toward the gravitational source. 
For $r=10\,GM/c^2$, for instance, the threshold for tidal acceleration is $V_{ta} \approx c/4$. If the interrogation point is instead at $r= 200 \, GM/c^2$, the threshold for tidal acceleration reduces to 
$V_{ta} \approx c/20$. The threshold in  case (i), i.e. $0 \le a \ll r$,  is roughly independent of the rotation of the source. On the other hand, the situation changes qualitatively for a very rapidly rotating Kerr source. There is a threshold in this case as well; however, for $V_0$ above this threshold, the tidal acceleration appears to be rather insignificant. For $r=10\,GM/c^2$, the threshold in case (ii) is $V_{ta} \approx 0.7\,c$. 

Finally, a comment is in order here regarding the fact that the Lorentz gamma factor can possibly  reach infinity in our limited approximation scheme. This singularity is clearly unphysical and will disappear in a more complete theory that includes all higher-order tidal effects. Such a treatment is beyond the scope of the present work, but exact Fermi coordinate systems have been constructed in the past to answer similar questions regarding the generalized Jacobi equation~\cite{CM3}. The trend of tidal acceleration toward $\Gamma \gg 1$, illustrated in the present paper, is encouraging, since it is consistent with the observational data regarding jets in AGNs. However, a proper comparison of theory with observation must incorporate this type of approach within a complete model for jets that includes electromagnetic effects as well. 

\appendix

\section{Fermi Coordinates}

Consider the world line of an arbitrary reference observer that is static in the exterior Kerr spacetime.  The observer has proper time $\tau$ and carries a non-rotating  tetrad frame $\lambda^{\mu}{}_{\hat \alpha} (\tau)$ along its path.  At each event $\bar{x}^\mu(\tau)$ along this world line, we imagine all spacelike geodesic curves that emanate orthogonally from this event and generate a local hypersurface. Let $x^\mu$ be an event on this hypersurface sufficiently close to the reference world line such that a \emph{unique} spacelike geodesic of proper length $\sigma$ connects it to $\bar{x}^\mu(\tau)$. We assign Fermi coordinates $X^\mu = (T, X^i)$  to event $x^\mu$, where
\begin{equation}\label{A1}
T := \tau\,, \qquad X^i := \sigma\, \xi^\mu(\tau)\, \lambda_{\mu}{}^{\hat i}(\tau)\,.
\end{equation}
Here,  $\xi^\mu(\tau)$, $\xi^\mu(\tau)\,\lambda_{\mu}{}^{\hat 0}(\tau) = 0$, is a unit spacelike vector  that is tangent to the unique geodesic connecting $\bar{x}^\mu(\tau)$ to $x^\mu$. That is, $\xi^\mu(\tau)\, \lambda_{\mu}{}^{\hat i}(\tau)$, for $i = 1, 2, 3$, are the corresponding direction cosines at proper time $\tau$ along the reference world line. The reference observer is permanently fixed at the spatial origin of the Fermi coordinate system. To simplify our notation, we have eliminated hats from the Fermi coordinate indices in this paper. 

The Fermi coordinate system is admissible in a cylindrical domain of radius $|\mathbf{X}| \sim \mathcal{R}$ in the spacetime around the reference world line. Here $\mathcal{R}$ is a certain minimal radius of curvature of spacetime along   $\bar{x}^\mu(\tau)$. 

The spacetime metric in Fermi coordinates is given by $-ds^2=\tilde{g}_{\mu \nu}\,dX^\mu \,dX^\nu$, where~\cite{mas77}
\begin{equation}\label{A2}
\tilde{g}_{00} = -1 -2A_i\,X^i - (A_i\,A_j +\tilde{R}_{0i0j})\,X^i\,X^j + O(|\mathbf{X}|^3)\,,
\end{equation}
\begin{equation}\label{A3}
\tilde{g}_{0i} = -\frac{2}{3} \,\tilde{R}_{0jik}\,X^j\,X^k + O(|\mathbf{X}|^3)\,,
\end{equation}
\begin{equation}\label{A4}
\tilde{g}_{ij} = \delta_{ij} -\frac{1}{3} \,\tilde{R}_{ikjl}\,X^k\,X^l + O(|\mathbf{X}|^3)\,.
\end{equation}
Here,  
\begin{equation}\label{A5}
A_i (T)  := \mathcal{A}_{\mu}\,\lambda^{\mu}{}_{\hat i}\,
\end{equation}
is the local acceleration of the reference observer and
\begin{equation}\label{A6}
\tilde{R}_{\alpha \beta \gamma \delta}(T) := R_{\mu \nu \rho \sigma}\,\lambda^{\mu}{}_{\hat \alpha}\,
\lambda^{\nu}{}_{\hat \beta}\,\lambda^{\rho}{}_{\hat \gamma}\,\lambda^{\sigma}{}_{\hat \delta}
\end{equation}
is the projection of the Riemann curvature tensor on the non-rotating tetrad frame of the reference observer. 

It is now straightforward to compute the Christoffel symbols in Fermi coordinates using Eqs.~\eqref{A2}--\eqref{A4}; in fact,  the non-zero components of the connection can be obtained from~\cite{mas77} 
\begin{eqnarray}
\tilde{\Gamma}^0{}_{00} &=& \frac{dA_i}{dT}\,X^i + O(|\mathbf{X}|^2)\,,
\nonumber\\
\tilde{\Gamma}^0{}_{0i} &=& A_i+(\tilde{R}_{0i0j}-A_iA_j)\,X^j + O(|\mathbf{X}|^2)\,,\nonumber\\
\tilde{\Gamma}^0{}_{ij} &=& \frac23\, \tilde{R}_{0(ij)k}\,X^k + O(|\mathbf{X}|^2)\,,\nonumber\\
\tilde{\Gamma}^i{}_{00} &=&  A_i+(\tilde{R}_{0i0j}+A_iA_j)\,X^j + O(|\mathbf{X}|^2)\,,\nonumber\\
\tilde{\Gamma}^i{}_{0j} &=& -\tilde{R}_{0kij}\,X^k+ O(|\mathbf{X}|^2)\,,\nonumber\\
\tilde{\Gamma}^i{}_{jk} &=& -\frac23\, \tilde{R}_{i(jk)l}\,X^l+ O(|\mathbf{X}|^2)\,.
\end{eqnarray}\label{A7}

\section{Acceleration and Curvature}

The gravitational source in our main jet Eqs.~\eqref{J2} and~\eqref{J3} is represented by the non-gravitational acceleration $A_1$ required to keep the reference observer static and  the curvature component $\tilde{R}_{0101}$. 

Let us first note that $A_i$, the projection of the 4-acceleration on $\lambda^{\mu}{}_{\hat i}$  can be directly calculated using Eqs.~\eqref{S5},~\eqref{F10} and~\eqref{F11}. For $\theta=0$, $\alpha=0$ and the only non-zero component of the acceleration is  then $A_1=\mathcal{A}_\mu \,e^{\mu}{}_{\hat 1}$ given in Eq.~\eqref{J1}. 

Next, the curvature of Kerr spacetime as measured by static observers using their natural tetrad system has been discussed in detail in Appendix B of Ref.~\cite{Bini:2016xqg}. Along the axis of rotation ($\theta=0, \pi$), we find that the tidal matrix is diagonal such that~\cite{Bini:2016xqg}
\beq\label{B1}
R_{\hat 0 \hat 1 \hat 0 \hat 1}=-2\, R_{\hat 0 \hat 2 \hat 0 \hat 2} = -2\,R_{\hat 0 \hat 3 \hat 0 \hat 3}= -\frac{2Mr (r^2-3a^2)}{(r^2+a^2)^3}\,.
\eeq
The non-rotating spatial frame $\lambda^{\mu}{}_{\hat i}$ is related to $e^{\mu}{}_{\hat i}$ by a simple rotation about the jet direction by Eq.~\eqref{F11}. It follows from the results of Ref.~\cite{Bini:2016xqg} that the same tidal matrix is obtained using the non-rotating frame; that is,
\beq\label{B2}
\tilde{R}_{0101}=-2\, \tilde{R}_{0202} = -2\,\tilde{R}_{0303}= -\frac{2Mr (r^2-3a^2)}{(r^2+a^2)^3}\,.
\eeq

Finally, for the jet in the southern hemisphere of the Kerr source, we note that $\theta = \pi$ and 
 $\lambda^{\mu}{}_{\hat 1} = - e^{\mu}{}_{\hat 1}$, so that $A_1 (\theta = \pi) = - A_1 (\theta =0)$. However, the curvature components will remain the same.  These facts together with Eq.~\eqref{J3a} are sufficient to demonstrate that the tidal dynamics of the two jet components are exactly the same.

\section{Integration of Eq.~\eqref{J9}}

Consider Eq.~\eqref{J9} and note that $dv/d\zeta=v\,dv/dz$. Let 
\beq\label{C1}
\chi=\frac12 v^2\,,
\eeq
so that Eq.~\eqref{J9} takes the form
\beq\label{C2}
\frac{d\chi}{dz}  +2\,(P z + Q)\,\chi +[- (q-\Phi \,p^2) z+ p]=0\,,
\eeq
where
\beq
\label{C3}
P :=2\,\Phi\, (q+\Phi \,p^2)\,, \qquad Q := -2\,\Phi\,p\,.
\eeq    
Let $\psi$ be an integrating factor such that
\beq
\label{C4}
\frac{d\psi}{dz}=2\,\psi  \,(P z + Q)\,;
\eeq                             
then, Eq.~\eqref{C2} can be expressed as 
\beq\label{C5}
\frac{d}{dz}(\chi\,\psi)+\psi\, [- (q-\Phi \,p^2) z+ p\,]=0\,.
\eeq      
From the solution of Eq.~\eqref{C4},
\beq\label{C6}
\psi(z)=C e^{P z^2 +2\,Q\,z}\,,
\eeq
where $C \ne 0$ is an integration constant, we can find the first integral of Eq.~\eqref{C5}. That is, 
integrating both sides of Eq.~\eqref{C5} from $0$ to $z$ leads to
\beq\label{C7}
\chi(z)\,\psi(z)-\chi(0)\,\psi(0)=-C\int_0^z   e^{Px^2 +2\,Q\, x}\,[- (q-\Phi \,p^2) x+ p\,] \,dx\,.
\eeq                       
Using the initial condition that at $z=0$, $v=v_0$, we find
\beq\label{C8}
v^2\,e^{P z^2 +2\,Q\,z}-v_0^2=-2\,\int_0^z   e^{Px^2 +2\,Q\, x}\,[- (q-\Phi \,p^2) x+ p\,]\, dx\,,
\eeq
or
\beq\label{C9}
v^2(z)= e^{-P z^2 -2\,Q\,z} \left\{v_0^2 -2\,\int_0^z   e^{Px^2 +2\,Q\, x}\,[- (q-\Phi \,p^2) x+ p\,]\, dx\right\}\,.
\eeq 
The integral in this expression can be expressed in terms of the error function. Finally, the path of the free test particle in the Fermi coordinate system is given by  
\beq\label{C10}
\zeta=\int_0^z \frac{dx}{v(x)}\,,
\eeq
assuming that $\zeta=0$ at $z=0$. 

It proves useful to define an effective ``potential energy" function $\mathcal{V}(z)$, 
\beq\label{C11}
\mathcal{V}(z)=2\,\int_0^z   e^{Px^2 +2\,Q\, x}\,[- (q-\Phi \,p^2) x+ p\,]\, dx\,.
\eeq
This function vanishes at  $z=0$, which is the location of the reference observer, and has an extremum at the rest point $z_0= p/(q-\Phi\,p^2)$ when  $q-\Phi\,p^2\ne 0$. It follows from Eq.~\eqref{C9} that the motion is confined to the region where
\beq\label{C12}
\mathcal{V}(z) \le v_0^2\,.
\eeq
To study certain general properties of this motion, we assume the Kerr source is such that $\Phi \ll 1$ and $0\le \eta <1$, so that $p>0$. It is then interesting to consider the cases where $q-\Phi\,p^2$ is positive, zero or negative. We denote the corresponding  $\mathcal{V}(z)$ functions by  $\mathcal{V}_{+}(z)$,  $\mathcal{V}_0(z)$ and  $\mathcal{V}_{-}(z)$, respectively.

%%%%%%%%%%%%%%%%%%%
%%%%%%%%%%%%%%%%%%%%
\begin{figure}
\begin{center}
\includegraphics[scale=0.75]{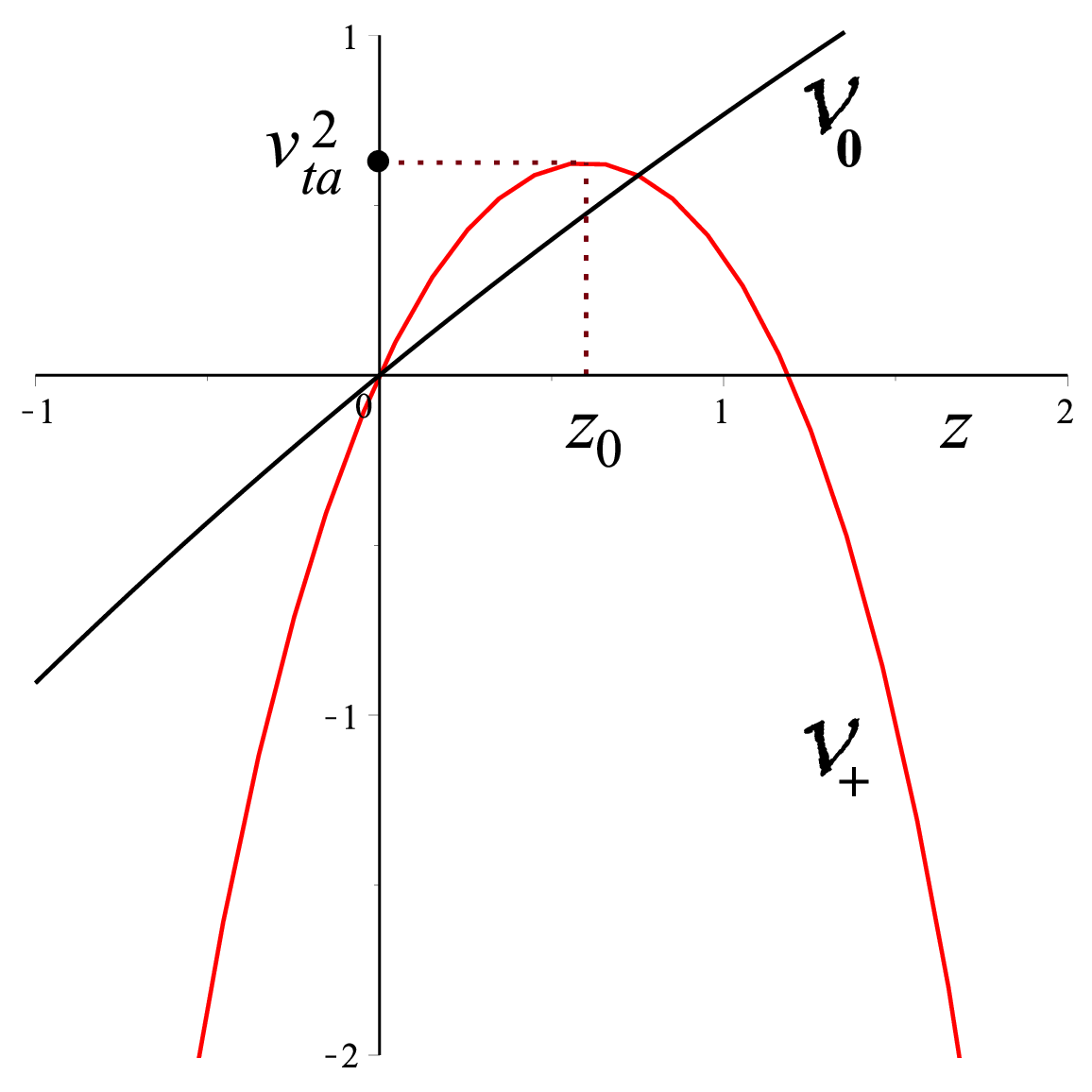} 
\caption{\label{fig.5} Plots of $\mathcal{V}_{+}$ and $\mathcal{V}_{0}$ versus $z$.  For $\mathcal{V}_{+}$, the parameters are $\Phi=1/10$ and $\eta=1/20$, so that $z_0 \approx 0.60$ and $v_{ta} \approx 0.79$, corresponding to the phase portrait in Figure 1. The parameters for $\mathcal{V}_{0}$ are  $\Phi=1/10$ and $\eta \approx 0.5715$. }
\end{center}
\end{figure}
%%%%%%%%%%%%%%%%%%%%
%%%%%%%%%%%%%%%%%%%%

If $q-\Phi\,p^2>0$, then the rest point $z_0=p/(q-\Phi\,p^2)$ is on the positive $z$ axis. For the sake of illustration, we plot in Figure 5 the function $\mathcal{V}_{+}(z)$, which is a potential barrier, for $\Phi=1/10$ and $\eta=1/20$. With this choice of parameters, $z_0 \approx 0.60$ and $v_{ta} \approx 0.79$. This case corresponds to the phase portrait presented in Figure 1. A particle moving along the positive $z$ axis and approaching the reference observer at $z=0$ with initial speed $v_0$, $0< v_0^2 \le v_{ta}^2$, where $v_{ta}^2 =\mathcal{V}_{+}(z_0)$, will encounter a turning point; that is, it will reach some $z$,  $0< z \le z_0$, reverse direction and fall back toward the source. On the other hand, if $v_0^2 > v_{ta}^2$, there is no turning point and the particle can get over the barrier and accelerate to $\Gamma=\infty$. This analysis is consistent with the phase portrait in Figure 1 and the fact that $v_{ta}$ is given by Eq.~\eqref{P8}.

If $q-\Phi\,p^2=0$, there is no rest point and $\mathcal{V}_{0}(z)$ increases monotonically with $z$. We plot   $\mathcal{V}_{0}(z)$ in Figure 5  for $\Phi=1/10$ and $\eta \approx 0.57$. A particle moving along the positive $z$ axis and approaching the reference observer at $z=0$ will reach some finite $z$, reverse direction, and fall back toward the source.

%%%%%%%%%%%%%%%%%%%
%%%%%%%%%%%%%%%%%%%%
\begin{figure}
\begin{center}
\includegraphics[scale=0.75]{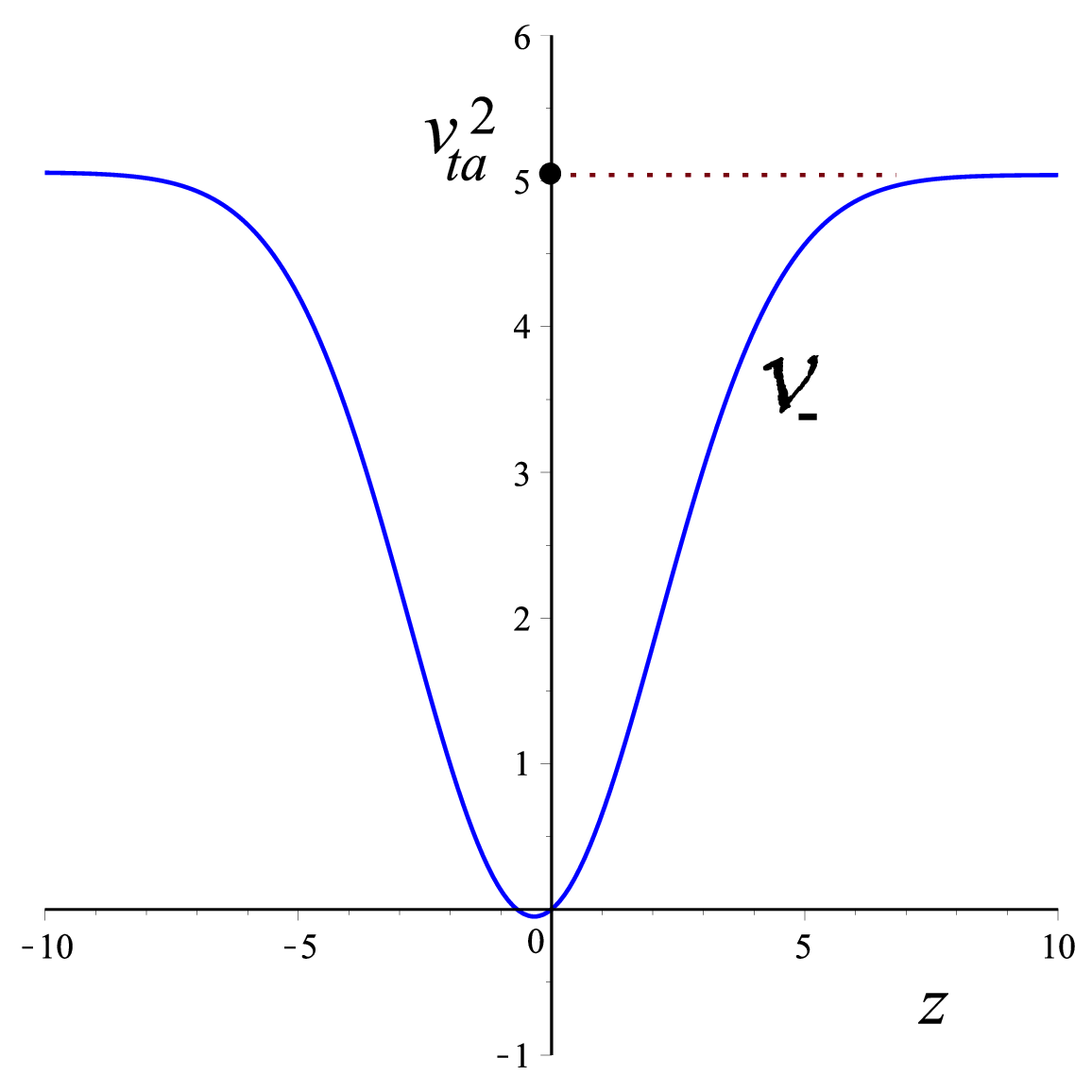}  
\caption{\label{fig.6} Plot of ${\mathcal V}_-$ versus $z$. The parameters are $\Phi=1/10$ and $\eta=8/10$ so that
$p\approx 0.14$, $q\approx -0.42$, $z_0 \approx-0.34$ and  $v_{ta}^2={\mathcal V}_-(\infty)\approx 5.04$. 
Note that ${\mathcal V}_-(-\infty)\approx 5.06>{\mathcal V}_-(\infty)$.}
\end{center}
\end{figure}
%%%%%%%%%%%%%%%%%%%%
%%%%%%%%%%%%%%%%%%%%

Finally, if $q-\Phi\,p^2 < 0$, then the rest point $z_0=p/(q-\Phi\,p^2)$ is on the negative $z$ axis. For the sake of illustration, we plot in Figure 6 the function $\mathcal{V}_{-}(z)$, which is a potential well, for $\Phi=1/10$ and $\eta=8/10$. With this choice of parameters, 
$\mathcal{V}_{-}(-\infty) > \mathcal{V}_{-}(\infty)$ and $v_{ta}^2 = \mathcal{V}_{-}(\infty)$ in this case, so that $z_0 \approx -0.34$ and $v_{ta} \approx 2.245$. This case corresponds to the phase portrait presented in Figure 4. If the initial conditions are such that $v_0^2 < v_{ta}^2$, then the motion is periodic and the particle oscillates repeatedly between the two turning points in the potential well. If, on the other hand,  $v_0^2 > v_{ta}^2$, then the motion is not confined and the  particle can, in principle, accelerate to $\Gamma=\infty$.

\section*{Acknowledgments}
D.B. thanks the Italian INFN (Salerno) for partial support.


\begin{thebibliography}{00}


\bibitem{Algaba:2016yvx} 
  J.~C.~Algaba, M.~Nakamura, K.~Asada and S.~S.~Lee,
  ``Resolving the Geometry of the Innermost Relativistic Jets in Active Galactic Nuclei'',
  Astrophys.\ J.\  {\bf 834},  65 (2017)
  %doi:10.3847/1538-4357/834/1/65
  [arXiv:1611.04075 [astro-ph.HE]].

\bibitem{Lee:2016ctn} 
  S.~S.~Lee, A.~P.~Lobanov, T.~P.~Krichbaum and J.~A.~Zensus,
  ``Acceleration of Compact Radio Jets on Sub-parsec Scales'',
  Astrophys.\ J.\  {\bf 826},  135 (2016)
 % doi:10.3847/0004-637X/826/2/135
  [arXiv:1604.02207 [astro-ph.GA]].

\bibitem{Walker:2016cic} 
  R.~C.~Walker, P.~E.~Hardee, F.~Davies, C.~Ly, W.~Junor, F.~Mertens and A.~Lobanov,
  ``Observations of the Structure and Dynamics of the Inner M87 Jet'',
  Galaxies {\bf 4}, 46 (2016)
  %doi:10.3390/galaxies4040046
  [arXiv:1610.08600 [astro-ph.HE]].
  
\bibitem{Mertens:2016rhi} 
  F.~Mertens, A.~P.~Lobanov, R.~C.~Walker and P.~E.~Hardee,
  ``Kinematics of the jet in M 87 on scales of 100--1000 Schwarzschild radii'',
  Astron.\ Astrophys.\  {\bf 595}, A54 (2016)
  %doi:10.1051/0004-6361/201628829
  [arXiv:1608.05063 [astro-ph.HE]].
   
\bibitem{Felice}
F. de Felice and O. Zanotti, 
``Jet dynamics in black hole physics: Acceleration during subparsec collimation'',
Gen. Relativ. Gravit. {\bf 32}, 1449 (2000).

\bibitem{BPu}
B. Punsly, \emph{Black Hole Gravitohydromagnetics} (Springer-Verlag, Berlin, 2008), 2nd ed.
  
\bibitem{CM1}
C.~Chicone and B.~Mashhoon, ``The generalized Jacobi equation", 
Classical Quantum Gravity {\bf 19}, 4231 (2002).

\bibitem{CM2}
C.~Chicone and B.~Mashhoon, ``Ultrarelativistic motion: Inertial and tidal effects in Fermi coordinates", Classical Quantum Gravity {\bf 22}, 195 (2005).

\bibitem{CM3}
C.~Chicone and B.~Mashhoon, ``Explicit Fermi coordinates and tidal dynamics in de Sitter and G\"odel spacetimes", Phys. Rev. D {\bf 74}, 064019 (2006).

\bibitem{Fender}
R.~Fender, 
``Relativistic outflows from X-ray binaries ('Microquasars')",
Lect. Notes Phys. {\bf 589}, 101 (2002). 

\bibitem{Fender:2014sia} 
  R.~Fender and E.~Gallo,
  ``An overview of jets and outflows in stellar mass black holes'',
  Space Sci.\ Rev.\  {\bf 183},  323 (2014)
  %doi:10.1007/s11214-014-0069-z
  [arXiv:1407.3674 [astro-ph.HE]].
  
  
\bibitem{Synge}
J.~L.~Synge, \emph{Relativity: The General Theory} (North-Holland, Amsterdam, 1971).  

\bibitem{Chicone:2010aa} 
  C.~Chicone and B.~Mashhoon,
  ``Gravitomagnetic Jets'',
  Phys.\ Rev.\ D {\bf 83}, 064013 (2011)
  %doi:10.1103/PhysRevD.83.064013
  [arXiv:1005.1420 [gr-qc]].
  
\bibitem{Chicone:2010hy} 
  C.~Chicone and B.~Mashhoon,
  ``Gravitomagnetic Accelerators'',
  Phys.\ Lett.\ A {\bf 375}, 957 (2011)
 % doi:10.1016/j.physleta.2010.12.076
  [arXiv:1005.2768 [gr-qc]].

\bibitem{Chicone:2010xr} 
  C.~Chicone, B.~Mashhoon and K.~Rosquist,
  ``Cosmic Jets'',
  Phys.\ Lett.\ A {\bf 375}, 1427 (2011)
  %doi:10.1016/j.physleta.2011.02.036
  [arXiv:1011.3477 [gr-qc]].

\bibitem{Chicone:2011ie} 
  C.~Chicone, B.~Mashhoon and K.~Rosquist,
  ``Double-Kasner Spacetime: Peculiar Velocities and Cosmic Jets'',
  Phys.\ Rev.\ D {\bf 83}, 124029 (2011)
  %doi:10.1103/PhysRevD.83.124029
  [arXiv:1104.5058 [gr-qc]].

\bibitem{Bini:2014esa} 
  D.~Bini and B.~Mashhoon,
  ``Peculiar velocities in dynamic spacetimes'',
  Phys.\ Rev.\ D {\bf 90},  024030 (2014)
%  doi:10.1103/PhysRevD.90.024030
  [arXiv:1405.4430 [gr-qc]].
  
  
\bibitem{Bini:2007zzb} 
  D.~Bini, F.~de Felice and A.~Geralico,
   ``Strains and axial outflows in the field of a rotating black hole'',
  Phys.\ Rev.\ D {\bf 76}, 047502 (2007)
  %doi:10.1103/PhysRevD.76.047502
  [arXiv:1408.4592 [gr-qc]].

\bibitem{JPGM}  
J.~Poirier  and G.~J.~Mathews,  
``Gravitomagnetic acceleration from black hole accretion disks",  
Classical Quantum Gravity {\bf 33}, 107001 (2016)
[arXiv:1504.02499 [gr-qc]].

\bibitem{Tucker:2016wvt} 
  R.~W.~Tucker and T.~J.~Walton,
  ``On Gravitational Chirality as the Genesis of Astrophysical Jets'',
  Classical Quantum Gravity  {\bf 34},  035005 (2017)
  %doi:10.1088/1361-6382/aa5325
  [arXiv:1609.07322 [gr-qc]].

\bibitem{Gariel:2007st} 
  J.~Gariel, M.~A.~H.~MacCallum, G.~Marcilhacy and N.~O.~Santos,
  ``Kerr Geodesics, the Penrose Process and Jet Collimation by a Black Hole'',
  Astron. Astrophys. {\bf 515}, A15 (2010)
  [gr-qc/0702123 [gr-qc]].

\bibitem{Gariel:2016vql} 
  J.~Gariel, N.~O.~Santos and A.~Wang,
  ``Observable acceleration of jets by a Kerr black hole'',
  Gen. Relativ. Gravit.  {\bf 49}, 43 (2017)
  %doi:10.1007/s10714-017-2208-9
  [arXiv:1610.01241 [gr-qc]].
 
\bibitem{Chandra}
S.~Chandrasekhar, \emph{The Mathematical Theory of Black Holes} (Clarendon, Oxford, 1983).
  
\bibitem{Bini:2016xqg} 
  D.~Bini and B.~Mashhoon,
  ``Relativistic gravity gradiometry,''
  Phys.\ Rev.\ D {\bf 94},  124009 (2016)
  %doi:10.1103/PhysRevD.94.124009
  [arXiv:1607.05473 [gr-qc]].
  
\bibitem{CH}
C. Chicone, {\it Ordinary Differential Equations with Applications}  (Springer-Verlag, New York, 2006), 2nd edn.

\bibitem{mas77}
B. Mashhoon,
``Tidal Radiation,''
Astrophys. J.\ {\bf 216}, 591-609 (1977).


\end{thebibliography}
\end{document}